# Oriented Attachment of ZnO Nanocrystals

*Dimitri Hapiuk[1], B. Masenelli[2,*], Karine Masenelli-Varlot[3], Dimitri Tainoff[1, †], Olivier Boisron[1], Clément Albin[1], Patrice Mélinon[1]*

[1]Université de Lyon, Lyon, F-69000, France and ILM, CNRS, UMR5306, Université Lyon 1, Villeurbanne, F-69622, France

[2]Université de Lyon, F-69000 Lyon, France and INL, CNRS, UMR 5270, INSA Lyon, F-69621 Villeurbanne, France

[3]Université de Lyon, F-69000 Lyon, France and MATEIS, CNRS, UMR 5510, INSA Lyon, F-69621 Villeurbanne, France

Abstract:

Self-organization of nanoparticles is a major issue to synthesize mesoscopic structures. Among the possible mechanisms leading to self-organization, the oriented attachment is efficient yet not completely understood. We investigate here the oriented attachment process of ZnO nanocrystals preformed in the gas phase. During the deposition in high vacuum, about 60% of the particles, which are uncapped, form larger crystals through oriented attachment. In the present conditions of deposition no selective direction for the oriented attachment is noticed. To probe the driving force of the oriented attachment, and more specifically the possible influence of the dipolar interaction between particles, we have deposited the same nanocrystals in the presence of a constant electric field. The expected effect was to enhance the fraction of domains resulting from the oriented attachment due to the increased interaction of the particle dipoles with the electric field. The multiscale analytical and statistical analysis (TEM coupled to XRD) shows no significant influence of the electric field on the organization of the particles. We therefore conclude that the dipolar interaction between nanocrystals is not the prominent driving force in the process. Consequently, we argue, in accordance with recent theoretical and experimental investigations, that the surface reduction, possibly driven by Coulombic interaction, may be the major mechanism for the oriented attachment process.

# 1. Introduction

The ability to synthesize mesostructures from nanoparticles as primary building blocks is a key issue of both fundamental and technological interests. One of the most prominent mechanisms involved in bottom-up strategies is the oriented attachment (OA) process.[1] It is a generic mechanism mainly observed in iono-covalent materials[2] such as $TiO_2$,[3,4] ZnO,[5,6,7,8] ZnS,[9,10] PbS,[11] PbSe,[12,13] $\alpha$-$Fe_2O_3$[14], $Sb_2S_3$[15] and YAG[16] but also in some metals[17] and even in natural[18] and biomineral systems.[19] Recent works have demonstrated the ability to synthesize large structures in a rather controlled way[20] such as mesoporous ellipsoids[8,21] and numerous nanorods or nanowires.[5,6,7,22]

However, the main driving force underlying the OA mechanism is still under debate. Two main hypotheses have been evoked. The first one is the interaction of dipoles carried by the nanoparticles. This is consistent with the fact that OA is mainly observed in iono-covalent materials even though it has also been reported for metallic particles.[17,23] The second main hypothesis is simply the reduction of the large surface energy of nanoparticles. Whereas direct measurements of particle dipoles of different origins have been made[24,25] and even harvested to self-organize nanoparticles,[26] no direct evidence of the role of the particle dipole in the OA process has been demonstrated. On the contrary, recent measurements[12] and first principle calculations on PbSe nanoparticles[13] go to show that the prominent role in the OA is played by the surface reduction.

In this work we investigate the mechanisms at stake in the OA. In order to probe the influence of the dipolar interaction between particles, we deposited iono-covalent nanocrystals in the presence or absence of an electric field (parallel or perpendicular to the substrate) and compare the formation of large crystalline domains resulting from the OA in both cases. Our strategy relies on a multiscale analysis by combining X-Ray Diffraction (XRD) to get crystalline information at the macroscopic scale and transmission electron microscopy (TEM) for local composition, crystallographic phase and orientation relations. If the dipolar interaction between particles is the prominent mechanism in the OA, the application of the external field during the deposition should enhance the mutual orientation of the particles and consequently their attachment. In order to maximize the hypothetical interaction of the external field with the particles, we have focused on ZnO nanoparticles. Beside its obvious major technological interest for opto-electronics,[27] photo-catalysis,[28] photovoltaics,[29] ZnO has the advantage of having a permanent electric dipole moment 11.4 times and 5.21 times larger than those of CdSe and CdS respectively[24,26] and should thus be highly sensitive to the influence of an electric field.

## 2. Methods

ZnO nanocrystals have been preformed in the gas phase by the LECBD technique (low energy cluster beam deposition) and subsequently deposited in a controlled environment (high vacuum). This procedure allows us to work with ligand-free stoichiometric nanocristals.[30] The synthesis technique used here has been detailed in previous work.[31] Briefly, a ZnO target made from a sintered powder (99.99% pure) is ablated by a pulse YAG:Nd laser (10 Hz repetition rate, 10 ns pulse duration). The ablation creates a plasma of Zn and O species which is first cooled by the continuous injection of a buffer gas at 20 mbar (see supporting information). The buffer gas is a mixture of 75% He and 25% $O_2$ to ensure the resulting clusters are stoichiometric.[30] While cooling the plasma, the buffer gas induces the formation of nucleation embryos (dimers and trimers). The plasma subsequently undergoes a supersonic adiabatic expansion while moving from the nucleation chamber at 20 mbar to the high vacuum deposition chamber at $10^{-7}$ mbar through a micrometric nozzle. During the expansion, the formation of clusters is achieved by accretion of embryos and atoms. The preformed clusters are then deposited on any given substrate without destruction since their kinetic energy per atom is several orders of magnitude lower than the binding energy per atom inside the clusters. Indeed, the velocity of the clusters is estimated to be about 500 m/s, corresponding to a kinetic energy per Zn or O atom of the order of 50 meV, which has to be compared to the covalent binding energy amounting to a few eV. Let us emphasize that up to now this technique performed to synthesize clusters of metallic and covalent materials has always led to nanostructured films in which the individual preformed clusters retain their integrity.[31] This behavior is referred to as the "memory effect".

For deposition in a static electric field parallel to the substrate, we have applied during deposition a 5 kV bias between two metallic electrodes (1 mm height) separated by 1 mm, creating thus a 5MV/m electric field, placed 0.8 mm above the substrate (TEM holey carbon grids with a thin film of amorphous carbon from Ted Pella inc.) and separated from it by an insulating spacer. For application of a field perpendicular to the substrate, the same set up has been used but with both electrodes connected to the 5 kV high voltage and the sample holder to the ground. Figure 1 shows both experimental set up configurations as well as the corresponding electric field maps. One can see a major difference between the two configurations: because of the conductive nature of the TEM grid, no planar electric field can exist at the sample surface. Therefore, the planar field is only operative in orienting the nanoparticle dipoles above the surface. On the contrary, when the electric field is applied perpendicularly to the substrate, its magnitude is maximal at the sample surface. Its effect is maximal and all the more efficient as the field is present all through the deposition.

The TEM analysis has been performed using a JEOL 2010F microscope equipped with a field emission gun operating at 200 keV. High resolution images were acquired with a Gatan CCD Orius camera and analyzed using Digital Micrograph software. Energy Dispersive Spectroscopy (EDS) spectra were collected with an Oxford Instruments INCA system. Size distribution was determined by acquiring bright field images of large zones of the sample, at random. The nanoparticle diameters were then measured on the images after segmentation on the grey level. To define the size and area of the crystalline domains in high resolution TEM images (HRTEM), we have considered that two particles in contact with identical reticular planes presenting an orientation mismatch of less than 5° were oriented and attached. The XRD analysis has been carried out on a Brucker D8 apparatus in the Bragg-Brentano configuration, using the Cu k$\alpha$ radiation.

**3. Evidence of oriented attachment**

Figure 2 (top) presents a TEM image of individual nanoparticles. The corresponding size distribution is a log-normal law peaked at 6 nm in diameter, typical of our accretion growth process. High resolution TEM images (right panels of figure 2) show that the clusters are crystallized in the wurtzite structure and *in situ* XPS analyses published[30] revealed that thin films resulting from the deposition of these particles were stoichiometric. Furthermore, energy dispersive spectroscopy (EDS) has shown that there is no significant composition fluctuation from one individual particle to another (see supporting information).

The wurtzite structure is further confirmed by XRD analysis on the cluster assembled films. However, as can be seen on Figure 3, the diffractogram contains two distinct contributions, a low magnitude one characterized by broad peaks and a high magnitude one made of sharp reflections corresponding to (10-10), (0002) and (10-11) diffracting planes. The weak contribution corresponds to clusters with a mean size of 6 ± 1 nm as deduced from the full-width at half-maximum (FWHM) using Debye-Scherrer's equation and assuming crystalline domains without defects. This mean size is in full agreement with the mode of the lognormal distribution determined from TEM micrographs. The same procedure applied to the intense sharp peaks gives a mean size of approximately 20 nm for the corresponding diffracting crystalline domains. Consequently, some clusters (64 ± 4% according to the ratio of the area of the sharp reflection peaks with the total area of the XRD signal) among the film have attached to form larger crystals and others have not. To confirm further this observation, we have co-deposited the same amount of clusters in a MgO matrix (generated by electron gun evaporation) to prevent the attachment. The deposition rates were chosen so that the individual ZnO clusters are separated by several tens of nanometers. The corresponding XRD diffractogram (cf. figure 3) is identical to the low magnitude contribution of the previous sample. Besides, the TEM analysis of this sample gives a log-normal law for the size dispersion (cf. figure 2 bottom left) with a mean diameter of 4.5 nm, in fair accordance with the value obtained for the isolated clusters considering that the contrast between the ZnO particles and the MgO matrix is rather poor. We thus conclude that some clusters among the ensemble interact to form larger particles and that this interaction occurs during the deposition (film formation) and not in the gas phase.

To get more insight into the formation of the larger particles we have analyzed by HRTEM thicker deposits in which the incident nanoparticles are no longer isolated. Figure 4 presents some selected micrographs. On these images the oriented attachment is clearly observed. In particular, in Figure 4 b) two distinct particles have attached perpendicularly to the [0001] axis and the resulting structure presents a bottleneck at the attachment location. A small mismatch of their respective lattices can be distinguished as well. In some cases the oriented attachment is not as perfect and leads to the formation of twin boundaries (cf. fig. 4f). This phenomenon has already been reported as a generation mechanism of crystalline defects (stacking faults, dislocations)[3]. The presence of large crystalline domains, consisting of several initial clusters, is clearly visible as the deposited amount of matter increases (see fig. 4a). Several examples are also given in fig. 5 and in the supporting information.

Contrary to few previous studies[5] and to recent numerical simulations[32], the OA in our case does not occur along specific directions. This absence of texture is clearly evidenced on the XRD diffractograms (cf. figure 3) where the sharp reflection peaks, corresponding to the large diffracting domains, present intensity ratios close to those of a powder. This is further evidenced by HRTEM in figure 5 and in the supporting information, where no specific orientation of the attachment axes is observed. For instance, in fig. 5b and c, we can distinguish two domains with their (10-10) planes in the Bragg diffraction orientation. However, the domain of fig. 5b is extended along these planes, suggesting an oriented attachment of other planes, whereas the domain of fig. 5c is extended along the (10-10) direction and thus seems to result from an OA of the (10-10) planes.

**4. Possible origin of the oriented attachment: influence of the dipolar interaction**

At this point, it is established that ZnO clusters, preformed in the gas phase, tend to gather through the OA mechanism during the deposition without any preferential attachment planes. Since the particles are passivated neither by capping ligands nor by air moisture (because of the deposition in high vacuum) we can rule out any chemical affinity between some capping shells or the surrounding medium. As evoked previously we are left mainly with two major hypotheses to explain the driving force of the OA mechanism. The first one is an interaction of dipolar origin and the second one is of chemical origin, namely the annihilation of facets.

The nanoparticle dipole can have three distinct causes. First, it can originate from the crystal lattice. This is the case of wurtzite structures with a non-ideal lattice parameter u/a ratio. The consequence is the existence of a cell dipole that scales as the volume of the nanoparticle, as reported for CdSe.[24,26] The second possible cause is the presence of polar facets. Polar facets can be obtained either naturally from the ionic crystal, such as {111} planes in the rocksalt structure or {0001} planes in the wurtzite structure or from non-polar facets missing one or few ions. The last possible origin of a dipole is the anisotropic shape of a neutral nanoparticle.[25] In all those cases, the dipole scales as the diameter of the particle. However, only dipoles originating from the cell or the {0001} facets, and consequently oriented along the c axis, can lead to oriented attachment since they are linked to the crystal network. The other dipoles are stochastically oriented with respect to the crystal network and are thus not expected to induce specific mutual orientation between particles. Hence they cannot lead to oriented attachment.

To check the possible influence of dipolar interactions, we have deposited the same nanoparticles in the presence of a static electric field, either parallel or perpendicular to the sample surface and compared their assembly to that of the previous nanoparticles deposited without any electric field. The value of the electric field E was set to 5 MV/m and 6.25 MV/m in the parallel and perpendicular geometries respectively so as to achieve an interaction between any dipole carried by the particles and the electric field larger than the thermal activation energy at room temperature. Indeed for a 6 nm diameter ZnO particle, the cell piezoelectric dipole D, which is proportional to the particle volume, amounts to 1700 Debye considering a piezoelectric spontaneous polarization[33,34] of 0.057 C/m$^2$ . The corresponding energy of interaction with the electric field is D.E, amounting to 0.16 eV (parallel geometry) or 0.2 eV (perpendicular geometry), a value much larger $k_B T$ at 300K (26 meV). On the other hand, the surface dipole S resulting from elementary charges located at opposite sites along the particle diameter would amount for the same particle to 283 Debye yielding an interaction energy of 30 meV (parallel geometry) or 37.5 meV (perpendicular geometry), a value slightly larger than $k_B T$ at room temperature. Therefore, if some dipoles of any origin are present among the particles, the application of the electric field should modify their mutual orientation and consequently their attachment probability. However, upon landing of the particles on the carbon film, the release of the kinetic energy may perturb the orientation presumably induced by the electric field.

We have plotted in figure 6 the fraction of diffracting particles, either individual or attached, for which we can observe a given {hkl} reticular plane family in diffraction condition. For the {10-11}, {10-10}, {10-12} and {10-13} plane families, the changes observed whether the electric field is applied perpendicularly or parallel to the sample surface are statistically significant (statistical analysis performed on more than 200 crystalline domains for each plan family, see supporting information for the details of the statistical analysis). Consequently, we can assert that the orientation of the field has an effect on the particle final orientation. Moreover, the variations are consistent with the expected ones. Indeed, when the electric field is perpendicular to the sample, since, as stated previously, the dipole interacting with the field is necessarily along the c-axis of the cell, we expect more particles with their c-axis parallel to the field and thus we should observe more planes parallel to this axis in diffraction condition. This is the case for the (10-10) planes. Conversely, all the others planes, intercepting the c-axis such as the (10-11), (10-12) and (10-13) planes, are more frequently observed when the field is parallel to the sample surface. This statement should also be true for the (0002) planes. However, the frequency of observation of these planes is identical for both orientations of the electric field. This may be due to the uncertainty peculiar to the (0002) planes in the assignment of the diffraction points in the FFT images. Anyway, the conclusion of this statistical analysis is that the electric field amplitude is large enough so as to modify the particle orientation.

The ultimate question is thus whether the electric field can enhance the OA process. To answer this question, a robust argument is to measure and compare the proportions of crystalline mono-domains among the particles observed in samples deposited with and without the electric field. The proportion of the area related to nanoparticles that are oriented and attached is 54%, 54.7% and 55% for the samples deposited without electric field and with the planar and perpendicular fields, respectively. No significant difference is observed from a statistical point of view. These values are also in fairly good agreement with the value of 64 ± 4% obtained from the XRD analysis mentioned previously. We can thus rule out the possible influence of a dipole carried by the incident clusters on the OA. This important statement is in accordance with recent experiments and simulations.[32] In particular, the work of Li and co-workers[20] emphasizes the effect of surface reduction in liquids rather than dipolar interaction by *in situ* TEM analysis. Also the simulations from Schapotschnikow *et al.*[13] on PbSe nanocrystals and from Zhang and Banfield on several nanocrystals[32] go to show that the prominent mechanism is surface reduction and not dipolar interaction.

If we sum up the previous observations, we have established that OA occurs during the deposition of uncapped ZnO nanoparticles preformed in the gas phase in high vacuum, on the substrate (not the gas phase). The OA process does not occur selectively along the [0001] direction as predicted for standard conditions [32] at thermodynamic equilibrium. Besides, since the application of an external electric field does not improve the OA efficiency, we would rather believe that the observed OA results from the annihilation of unstable surfaces of adjacent particles. To explain these statements, we can consider a layer of ZnO nanocrystals deposited on a substrate (as depicted in figure 7) onto which other nanoparticles are subsequently impinging during the deposition. The deposition is obviously a process where the impinging nanoparticles are not free to move, rotate and choose the best orientation to attach (no thermodynamic equilibrium). However, when a subsequent cluster hits the already present nanoparticles, it transfers some kinetic energy which allows the deposited clusters to move and rotate for a very short period. During this time, the high Coulombic interaction between clusters (all the higher as they are at close range) can lead to rearrangement and to the oriented attachment of the two clusters if they are not initially too disoriented. In other words, we suspect the subsequent impinging clusters and buffer gas molecules to provide some extra energy and thus some degrees of freedom to the already present clusters; a freedom of motion which has been pointed out by Li and co-workers[20] in their study in liquids as a key parameter in the OA process.

## 5. Conclusions

In summary, we have studied the organization of ZnO nanocrystals preformed in the gas phase. We have observed that during the deposition in high vacuum about 60% of the particles, which have uncapped surfaces, form larger crystals due to the OA process. In our conditions of deposition, no selective direction for the OA is noticed. To probe the driving force of the OA, and more precisely the possible influence of the dipolar interaction between particles, we have deposited the same nanocrystals in the presence of a constant electric field. The expected effect was to enhance the fraction of domains resulting from the OA due to the increased interaction between the particle dipoles and the electric field. Despite the proof that the electric field orientation controls the particle orientation, no significant effect has been observed. We thus conclude and confirm in accordance with recent theoretical and experimental investigations on other iono-covalent systems that the dipolar interaction between nanocrystals is probably not a prominent driving force in the OA process. Rather, the surface reduction, possibly driven by Coulomb interaction, is the main mechanism for the OA. This statement is a major fact in order to control the growth of super-architectures and mesostructures from initial nanocrystals.

**Funding Sources**

This work was supported by the INSA Lyon BQR project 2012. KMV acknowledges the Institut Universitaire de France (IUF) for financial support.

ACKNOWLEDGMENT

The authors acknowledge the PLYRA (plateforme lyonnaise de recherche sur les agrégats, http://www-lpmcn.univ-lyon1.fr/plyra/) facility for access to the nanoparticle generators. Thanks are due to the CLYM (Centre Lyonnais de Microscopie http://www.clym.fr) for the access to the microscope JEOL 2010F. The authors also acknowledge the help of Qian Rong in the analysis of TEM clichés.

**Supporting Information**

LECBD set up; EDS local analysis of the ZnO nanocrystals; HRTEM images of large as-deposited nanoparticles; presence of large crystalline domains resulting from OA and their mutual orientation and extension; statistical analysis of the effect of the orientation of the electric field on the particle orientation; dynamics of a particle cluster observed by TEM. This material is available free of charge *via* the Internet at http://pubs.acs.org.

Corresponding Author

* To whom correspondence should be addressed: tel. +33 4 72437472. E-mail: bruno.masenelli@insa-lyon.fr

† Present address : Institut Néel, CNRS/UJF UPR2940, 25 rue des martyrs BP166, 38042 Grenoble Cedex 9

**Figure captions :**

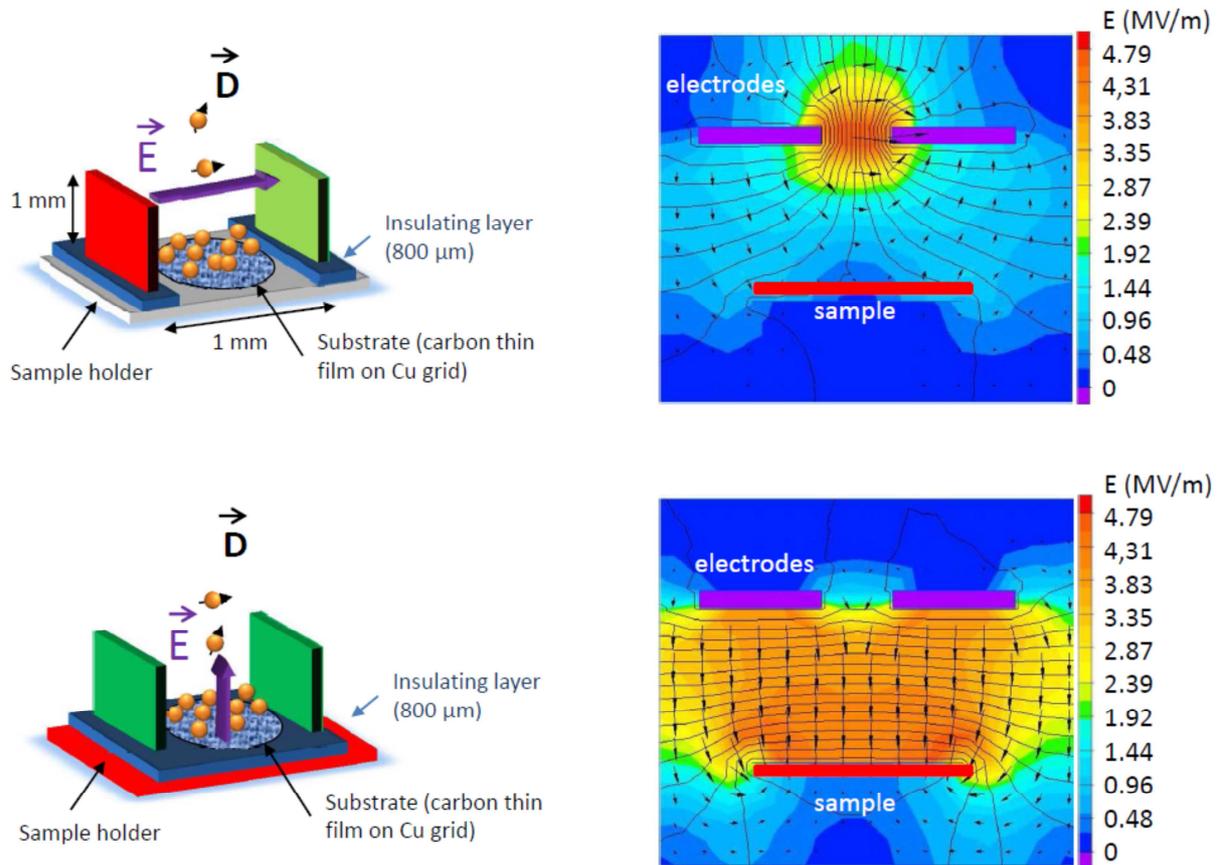

Figure 1: scheme of the experimental set up used to apply the static electric field during the deposition of the clusters. The field is constant, amounting to 5 MV/m (a 5kV bias across a 1 mm gap). D stands for the dipole carried by the nanoparticles. Two configurations with the electric field parallel or perpendicular to the sample have been used. The corresponding maps of field intensity are presented on the right part.

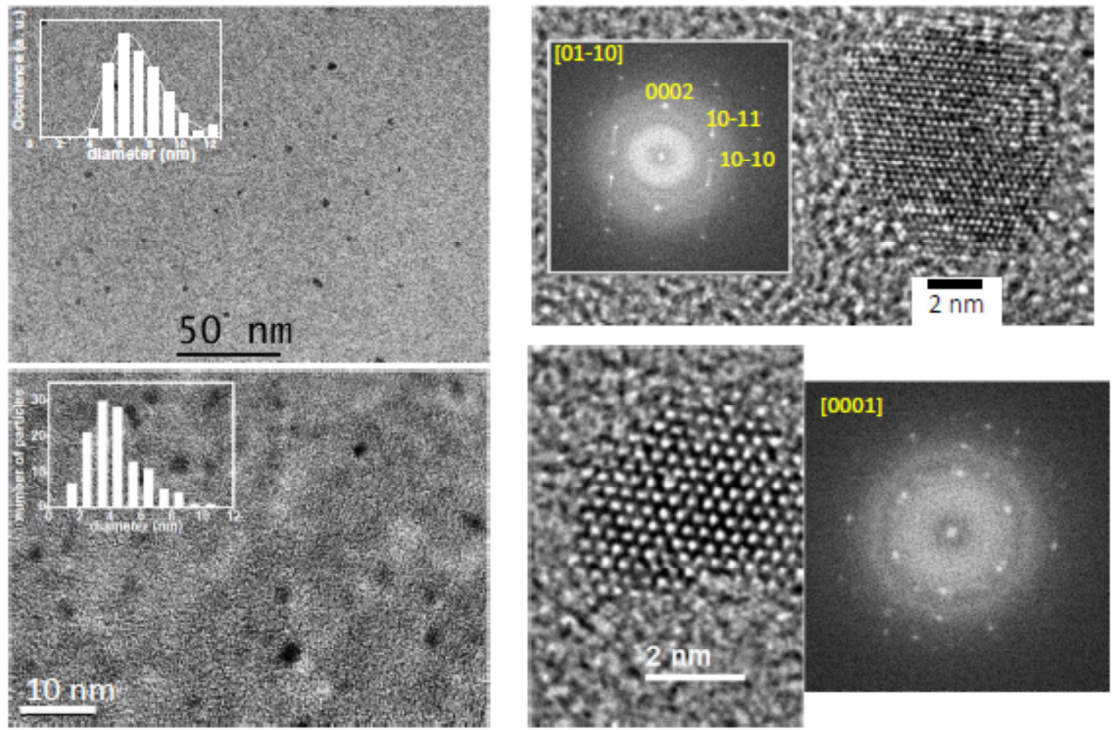

Figure 2: Top left: bright field TEM image of the as-deposited ZnO clusters and the corresponding size distribution function (inset). Bottom left: bright TEM image and size distribution function (inset) of the ZnO clusters dispersed in MgO matrix. Top and bottom right: two examples of HRTEM images of individual clusters. The particles are perfectly crystallized in the wurtzite structure, as evidenced by the FFT of the images. The particles of the top and bottom right panels have their zone axis oriented along the [01-10] and [0001] directions respectively.

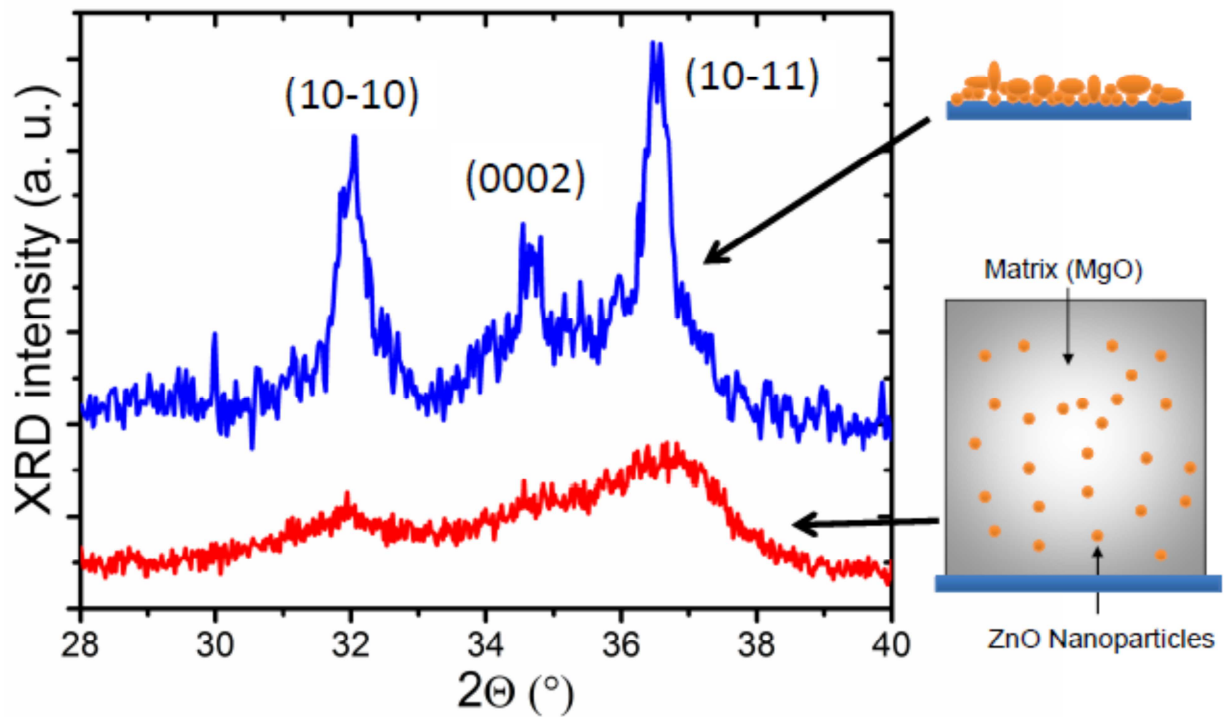

Figure 3: XRD spectra of the ZnO clusters assembled film (top) and of the same clusters dispersed in a MgO matrix (bottom). The spectrum of the cluster assembled film presents two distinct contributions, one identical to that of the clusters dispersed in MgO with broad peaks characteristic of small crystalline domains and a second one with sharp peaks indicative of large crystalline domains. The sharp contribution amounts to 64% of the overall spectrum.

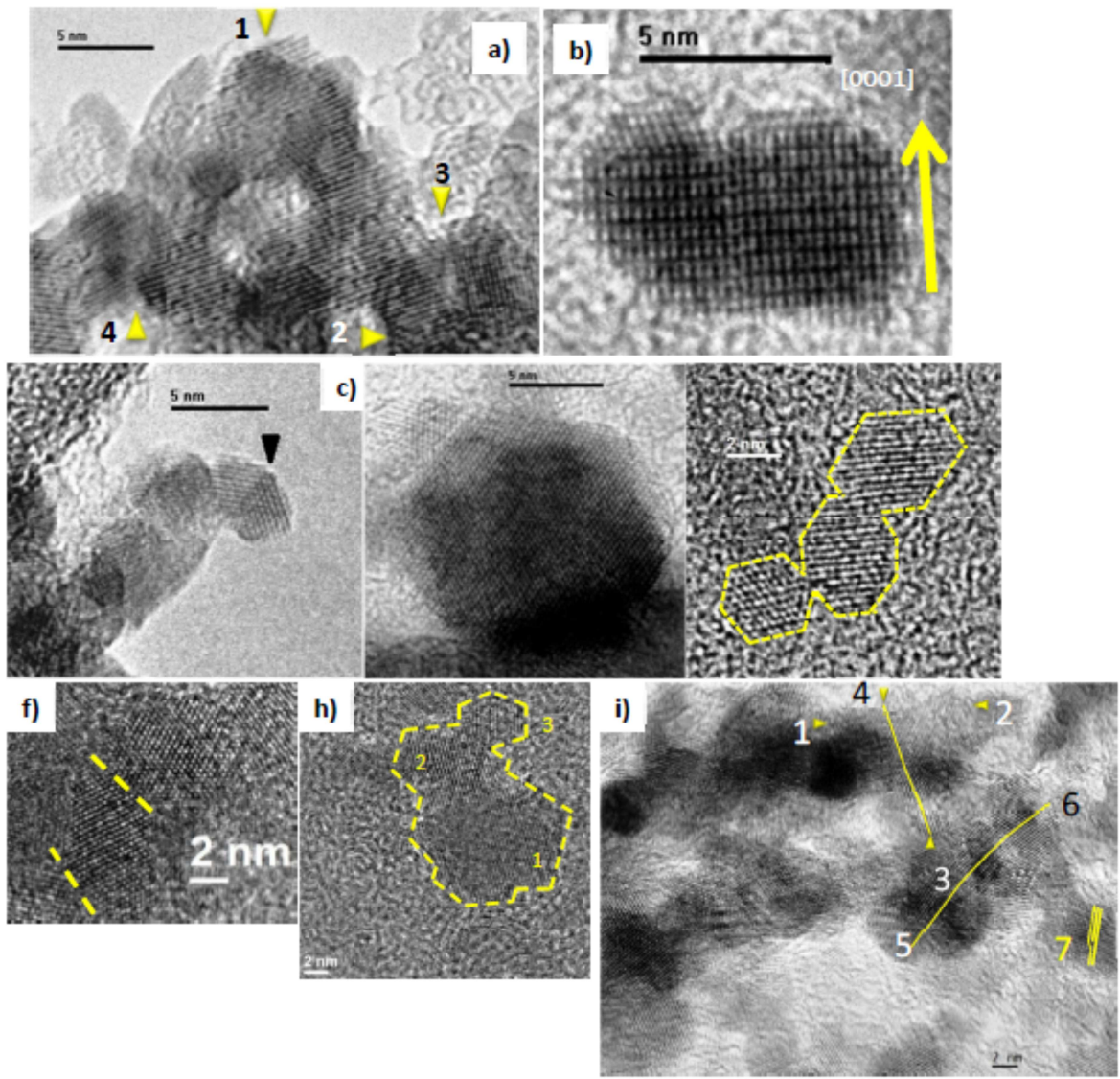

Figure 4: a) a typical large crystalline domain resulting from the OA of several initial small clusters. The arrows labeled with numbers 1 to 4 precise the extension of the domain; b) crystalline domain resulting from the OA of two distinct clusters. The [0001] direction is indicated by the arrow. A small mismatch of the reticular planes can be seen; c) two attached particles. The arrow highlights the attachment plane; d) other example of several particles attached to form a larger faceted domain. On the top left part of the image, a small particle attached to the large cluster can be distinguished; e) three attached particles. The facets of the particles are indicated by dotted lines; f) Two examples of twin boundaries resulting from the oriented attachment of ZnO nanocrystals. The yellow dotted lines indicate the locations of the twin boundaries; h) several particles forming a large crystalline domain. The fact that the domain results from the attachment of several particles is evidenced by the contour of the domain. Note that the particle labeled 3 is not in perfect match with particle 2. i) HRTEM image of several crystalline domains resulting from the OA: a large domain spans from point 1 to point 2 and from point 3 to point 4. The fact that it results from the attachment of several particles is observable by the slight variations of the yellow line running from point 3 to point 4. The same statement holds for the domain spanning from point 5 to 6. Eventually, in the particle labeled by point 7, the mismatch of the oriented attachment has resulted in an edge dislocation.

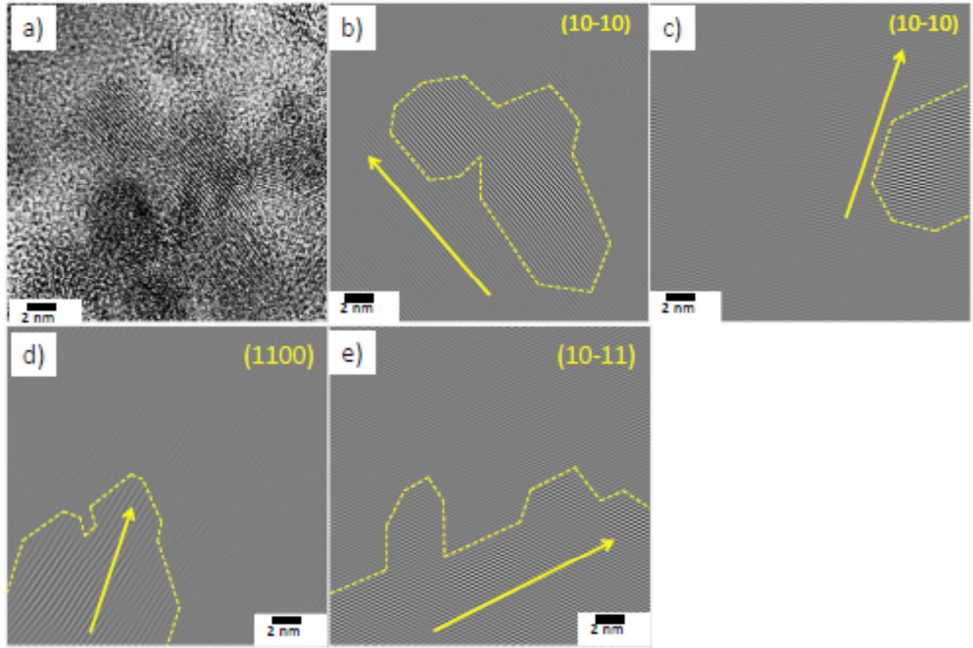

Figure 5: a) HRTEM image of several crystalline domains resulting from the OA (top left). The distinct domains are visible on the images reconstructed from the filtered FFT image ( b) to e)). The arrow highlights the direction of the domain extension. The crystalline domain in b) is extended along the (10-10) diffracting planes. On the contrary, the domain in c) is extended perpendicularly to these planes, suggesting that the OA is not selectively oriented in our experiments; d) and e) are domains with their (1100) and (10-11) planes in the Bragg diffraction condition. In d) the domain extension is perpendicular to the (1100) planes while it is perpendicular to the [10-11] direction in e).

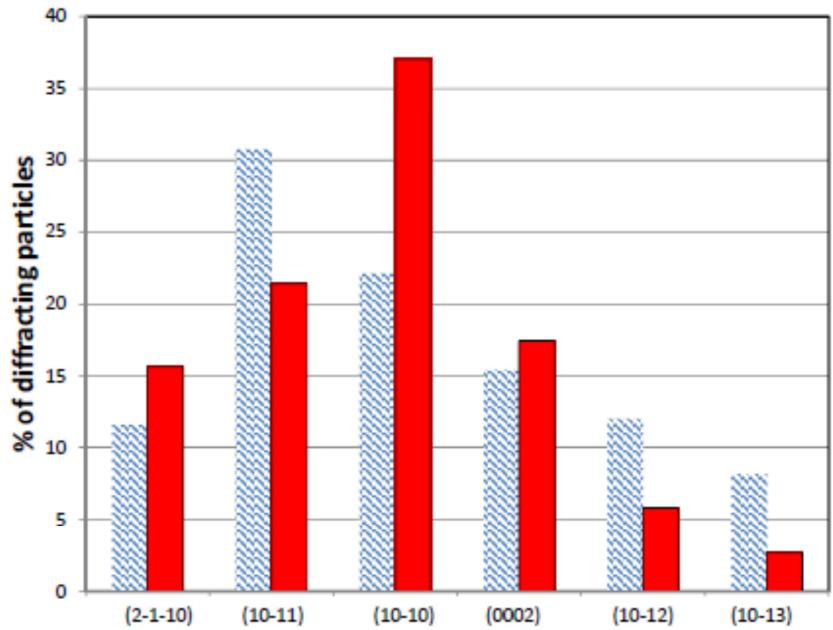

Figure 6: effect of the electric field orientation on the particle orientation: fraction of diffracting particles, either individual or attached, for which we can observe a given {hkl} reticular plane family in diffraction condition. Full bars: field perpendicular to the sample, striped bars: field parallel to the sample.

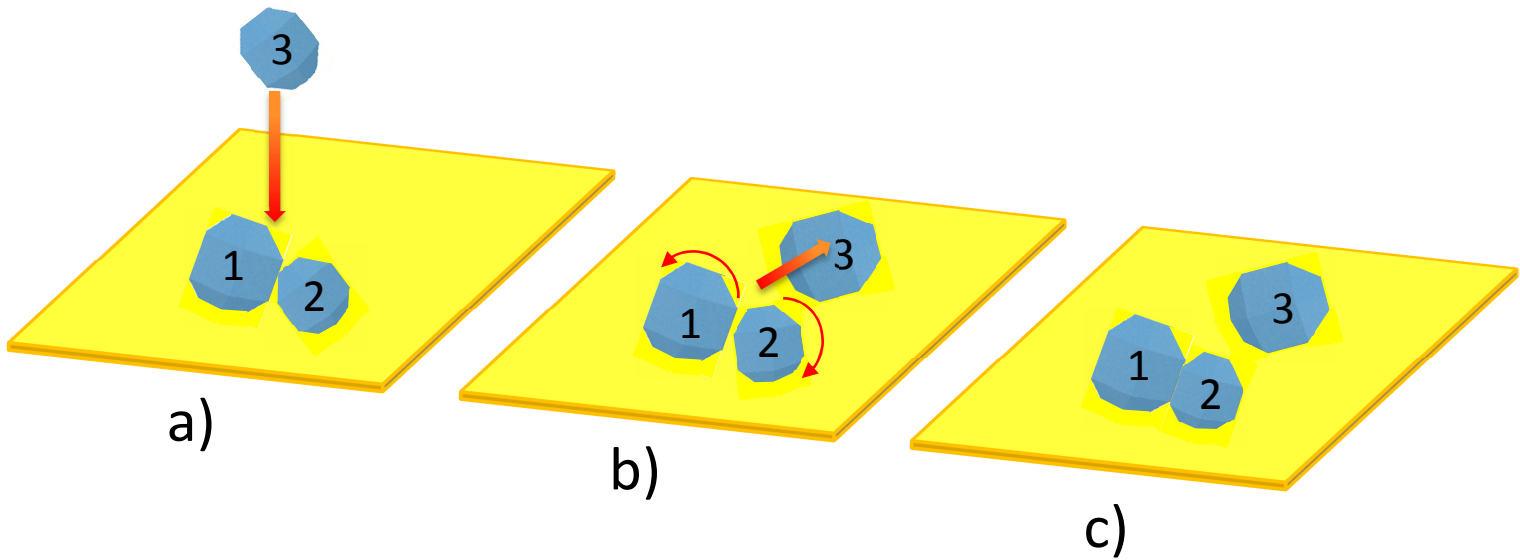

Figure 7: proposed scheme of the OA process occurring during the deposition of clusters preformed in the gas phase: a) particles 1 and 2 are already deposited, touching but not oriented. Particle 3 is arriving from the cluster beam. b) Upon impinging, particle 3 releases some kinetic energy to the other particles. This energy excess sets them free from the substrate interaction and their mutual interaction. Particles 1 and 2 are free to move and explore several configurations of orientation for a short period of time. If, during this period, they find an orientation of attachment, leading to the reduction of surface, OA will be promoted. c) Particles 1 and 2 are now oriented and attached while particle 3 has landed on the substrate.

[1]     Yin, Y.; Alivisatos, A.P. Colloidal Nanocrystal Synthesis and the Organic-Inorganic Interface *Nature* **2005**, *437*, 664-670

[2]     Li, Z.; Xu, F.; Sun, X.; Zhang, W. Oriented Attachment in Vapor: Formation of ZnO Three-Dimensional Structures by Intergrowth of ZnO Microcrystals *Crystal Growth & Design* **2008**, *8*, 805-807

[3]     Penn, R.L.; Banfield, J.F. Imperfect Oriented Attachment: Dislocation Generation in Defect-Free Nanocrystals *Science* **1998**, *281*, 969-971

[4]     Penn, R.L.; Banfield, J.F. Morphology Development and Crystal Growth in Nanocrystalline Aggregates under Hydrothermal Conditions: Insights from Titania *Geoch. Cosm. Acta* **1999**, *63*, 1549-1557

[5]     Pacholski, C.; Kornowski, A.; Weller, H. Self-Assembly of ZnO: From Nanodots to Nanorods *Angew. Chem. Int. Ed.* **2002**, *41*, 1188-1191

[6]     Liu, X.; Jin, Z.; Liu, Z.; Yu, K.; Bu, S. Nanostructured ZnO Flms Obtained by a Basic Erosion Method *Appl. Surf. Sci.* **2006**, *252*, 8668-8672

[7]     Zhang, D.-F.; Sun, L.-D.; Yin, J.-L.; Yan, C.-H.; Wang, R.-M., Attachment-driven Morphology Evolvement of Rectangular ZnO Nanowires. *J. Phys. Chem. B* **2005**, *109*, 8786-8790

[8]     Liu, Y.;  Shi, J.; Peng, Q.; Li, Y. Self-Assembly of ZnO Nanocrystals Into Nanoporous Pyramids: High Selective Adsorption and Photocatalytic Activity**.** *J. Mater. Chem.* **2012**, *22*, 6539-6541

 [9]     Huang, F.; Gilbert, B.; Zhang, H.; Banfield, J.F. Reversible, Surface-Controlled Structure Transformation in Nanoparticles Induced by an Aggregation State *Phys. Rev. Lett.* **2004**, *92*, 155501

[10]     Fang, X.S.; Zhai, T.Y.; Gautam, U.K.; L, L.; Wu, L.M.; Bando, Y.; Golberg, D. ZnS Nanostructures: From Synthesis to Applications, *Prog. Mater. Sci.* **2011**, *56*, 175-287

[11]     Schliehe, C.; Juarez, B.H.; Pelletier, M.; Jander, S.; Greshnykh, D.; Nagel, M.; Meyer, A.; Foerster, S.; Kornowski, A.; Klinke, C. *et al.* Ultrathin PbS Sheets by Two-Dimensional Oriented Attachment *Science* **2010**, *329*, 550-553

[12]     Van Huis, M.A.; Kunneman, L.T.; Overgaag, K.; Xu, Q.; Pandraud, G.; Zandbergen, H.W. Vanmaekelberg, D. Low-Temperature Nanocrystal Unification Through Rotations and Relaxations Probed by in-situ Transmission Electron Microscopy *Nano Lett*. **2008**, *8*, 3959-3963

[13]     Schapotschnikow, P.; Van Huis, M.A.; Zandbergen, H.W.; Vanmaekelbergh, D.; Vlugt, T.J.H. Morphological Transformations and Fusion of PbSe Nanocrystals Studied Using Atomistic

Simulations *Nano Lett*. **2010**, *10*, 3966-3971

[14]     Frandsen, C.; Bahl, C.R.H.; Lebech, B.; Lefmann, K.; Kuhn, L.T.; Keller, L.; Andersen, N.H.; von Zimmermann, M.; Johnson, E.; Klausen S.N. *et al.* Oriented Attachment and Exchange Coupling of $\alpha$-$Fe_2O_3$ Nanoparticles *Phys. Rev. B* **2005**, *72*, 214406

[15] Lu, Q.; Zeng, H.; Wang, Z.; Cao, X.; Zhang, L. Design of $Sb_2O_3$ Nanorod-Bundles: Imperfect Oriented Attachment. *Nanotechnology* **2006**, *17*, 2098-2104

[16]     Jia, N.; Zhang, X.; He, W.; Hu, W.; Meng, X.; Du, Y.; Jiang, J.; Du, Y. Property of YAG: Ce Phosphors Powder Prepared by Mixed Solvothermal Method *J. All. Comp.* **2011**, *509*, 1848-1853

[17] Halder, A.; Ravishankar, N. Ultrafine Single-Crystalline Gold Nanowire Arrays by Oriented Attachment *Adv. Mater.* **2007**,*19*,1854-1858

[18] Banfield, J.F.; Welcj, S.A.; Zang, H.; Ebert, T.T.; Penn, R.L. Aggregation-Based Crystal Growth and Microstructure Development in Natural Iron Oxyhydroxide Biomineralization Products *Science* **2000**, *289*, 751-754

[19] Killian, C.E.; Metzler, R.A.; Gong, Y.U.T.; Olson, I.C.; Aizenberg, J.; Politi, Y.; Wilt, F.H.; Scholl, A.; Young, A.; Doran, A. *et al.* Mechanism of Calcite Co-Orientation in the Sea Urchin Tooth *J. Am. Chem. Soc.* **2009**, *131*, 18404-18409

[20] Li, D.; Nielsen, M.H.; Lee, J.R.I.; Frandsen, C.; Banfield, J.F.; De Yoreo, J.J. Direction-Specific Interactions Control Crystal Growth by Oriented Attachment *Science* **2012**, *336*, 1014-1018

[21] Liu, Y.; Wang, D.; Peng, Q.; Chu, D.; Liu, X.; Li, Y. Directly Assembling Ligand-Free ZnO Nanocrystals into Three-Dimensional Mesoporous Structures by Oriented Attachment *Inorg. Chem.* **2011**, 50 5841-5847

[22] Yang, M.; Pang, G.; Li, J.; Jiang, L.; Feng, S. Preparation of ZnO Nanowires in a Neutral Aqueous System: Concentration Effect on the Orientation Attachment Process *Eur. J. Inorg. Chem.* **2006**, *19*, 3818-3822

[23] Yang, S.K.; Cai, W.P.; Liu, G.Q.; Zeng, H.B. From Nanoparticles to Nanoplates: Preferential Oriented Connection of Ag Colloids during Electrophoretic Deposition *J. Phys. Chem. C* **2009**, *113*, 7692-7696

[24] Li, L-S. ; Alivisatos, A.P. Origin and Scaling of the Permanent Dipole Moment in CdSe Nanorods *Phys. Rev. Lett.* **2003**, *90*, 097402

[25]     Shim, M.; Guyot-Sionnest, P. Permanent Dipole Moment and Charges in Colloidal Semiconductor Quantum Dots  *J. Chem. Phys.* **1999**, *111*, 6955-6964

[26]     Ryan, K.M.; Mastroianni, A.; Stancil, K.A.; Liu, H.; Alivisatos, A.P. Electric-Field-Assisted Assembly of Perpendicularly Oriented Nanorod Superlattices *Nanolett*. **2006**,*6*, 1479-1482

[27]     Tsukazaki, A.; Ohtomo, A.; Onuma, T.; Ohtani, M.; Makino, T.; Sumiya, M.; Ohtani, K.; Chichibu, S.F.; Fuke, S.; Segawa, Y. *et al.* Repeated Temperature Modulation Epitaxy for p-Type Doping and Light-Emitting Diode Based on ZnO *Nat. Mat.* **2005**, 4, 42-46

[28]     Hariharan, C. Photocatalytic Degradation of Organic Contaminants in Water by ZnO Nanoparticles: Revisited *Appl. Catal. A* **2006**, *304*, 55-61

[29]     Oosterhout, S.D.; Wienk, M.M.; van Bavel, S.S.; Thiedmann, R.; Koster, L.J.A; Gilot, J.; Loos, J.; Schmidt, V.; Janssen, R.A.J. The Effect of Three-Dimensional Morphology on the Efficiency of Hybrid Polymer Solar Cells *Nat. Mat.* **2009**, *8*, 818-824

[30]     Tainoff, D.; Masenelli, B.; Boisron, O.; Guiraud, G.; Melinon, P. Crystallinity, Stoichiometry and Luminescence of High Quality ZnO Nanoclusters *J. Phys. Chem. C* **2008,** *112*, 12623-12627

[31]     Perez, A.; Melinon, P.; Dupuis, V.; Bardotti, L.; Masenelli, B.; Tournus, F.; Prevel, B.; Tuaillon-Combes J., Bernstein, E.; Tamion, A. *et al.* Functional Nanostructures from Clusters *Int. J. Nanotechnol*. **2010**, *7*, 523-574

[32]     Zhang, H.; Banfield, J. F. Energy Calculations Predict Nanoparticle Attachment Orientations and Asymmetric Crystal Formation *J. Phys. Chem. Lett*. **2012**, *3*, 2882-2886

[33] Bernardini, F.; Fiorentini, V.; Vanderbilt, D. Spontaneous Polarization and Piezoelectric Constants of III-V Nitrides *Phys. Rev. B* **1997**, *56*, R10024-R10026

[34]    Noel, Y.; Zicovich-Wilson, C.M.; Civalleri, B.; D'Arco, Ph.; Dovesi, R. Polarization Properties of ZnO and BeO:    An *ab initio* Study Through the Berry Phase and Wannier Functions Approaches *Phys. Rev. B* **2001**, *65*, 014111